# Glycerol confined in zeolitic imidazolate frameworks: The temperature-dependent cooperativity length scale of glassy freezing

M. Uhl,[1] J. K. H. Fischer,[1] P. Sippel, H. Bunzen,[2] P. Lunkenheimer,[1,a] D. Volkmer,[2] and A. Loidl[1]

[1]Experimental Physics V, Center for Electronic Correlations and Magnetism, University of Augsburg, 86135 Augsburg, Germany
[2]Chair of Solid State and Materials Chemistry, Institute of Physics, University of Augsburg, 86135 Augsburg, Germany

[a]Electronic mail: peter.lunkenheimer@physik.uni-augsburg.de

In the present work, we employ broadband dielectric spectroscopy to study the molecular dynamics of the prototypical glass former glycerol confined in two microporous zeolitic imidazolate frameworks (ZIF-8 and ZIF-11) with well-defined pore diameters of 1.16 and 1.46 nm, respectively. The spectra reveal information on the modified $\alpha$ relaxation of the confined supercooled liquid, whose temperature dependence exhibits clear deviations from the typical super-Arrhenius temperature dependence of the bulk material, depending on temperature and pore size. This allows assigning well-defined cooperativity length scales of molecular motion to certain temperatures above the glass transition. We relate these and previous results on glycerol confined in other host systems to the temperature-dependent length scale deduced from nonlinear dielectric measurements. The combined experimental data can be consistently described by a critical divergence of this correlation length as expected within theoretical approaches assuming that the glass transition is due to an underlying phase transition.

## I. INTRODUCTION

Glasses and glass-forming materials are essential components of everyday life. Although the glass transition has been intensely studied for many decades, it is still not fully understood. The mystery of glassy freezing has not been completely unraveled and its investigation remains an important tasks of modern solid state physics.[1,2,3,4] In most cases, the temperature evolution of the molecular dynamics of glass-forming liquids does not simply follow a thermally activated (Arrhenius) behavior but rather more strongly slows down when approaching the glass-transition temperature $T_g$. Understanding the molecular origin of this non-canonical behavior represents a great scientific challenge. Often, an increasingly cooperative motion of the glass-forming entities (molecules, ions, polymer segments, etc.) under cooling is presumed to explain the observed super-Arrhenius behavior.[2,5,6,7,8] This notion is based on theories assuming that the glass transition is caused by an underlying phase transition into a state with a specific kind of "amorphous order".[5,6,9] This would imply the divergence of a characteristic correlation length $L_{corr}$, characterizing the cooperatively rearranging regions, at a temperature significantly below $T_g$. However, this transition cannot be directly detected because the supercooled liquid falls out of equilibrium when cooling below the glass temperature. Nevertheless, this phase-transition scenario was recently nicely corroborated by nonlinear dielectric measurements from which the relative temperature dependence of $L_{corr}$ was deduced, albeit not providing absolute values.[10,11,12,13] Other very useful attempts to learn more about the molecular dynamics and the length scales involved in the glass transition are investigations of supercooled liquids that are confined in spaces of nanometer size.[14,15,16,17,18,19,20,21,22,23,24,25,26] In bulk glass formers, the size of the cooperatively rearranging regions is supposed to increase with decreasing temperature, which should lead to an increase of the effective energy barriers and, consequently, to a stronger temperature dependence of relaxation times than in the Arrhenius case. But when the glass former is confined in a pore of defined geometry, below a certain temperature the growing correlation length $L_{corr}$ of the cooperative regions will exceed the pore diameter and, ideally, a crossover from non-Arrhenius to Arrhenius behavior would be expected.

Many different materials have been reported as host materials for measurements in confined geometries (for a thorough overview, see Ref. 23). These can be classified as materials with ordered pores or with disordered pores. Furthermore, the geometry of the pores is also important for the properties of the confined liquid[19] and, thus, it is convenient to differentiate between 3D- (pores), 2D- (layers), and 1D-confinement (channels). Generally, for a meaningful comparison with bulk properties a 3D-confinement seems preferable, thus avoiding dimensionality effects obscuring the cooperativity-related confinement effects. In previous works, confinement measurements on glass formers have been studied for various host materials of different pore sizes.[18,19,20,22,23] We have recently reported[25] that metal-organic frameworks (MOFs) are well suited 3D host systems for confinement investigations.

In general, MOFs represent three-dimensional networks composed of metal ions or clusters and organic linkers, thereby forming extended crystalline frameworks with significant porosity and three-dimensional pore geometry.[27,28,29] Due to their highly ordered structure and design versatility, MOFs with various, well-defined pore sizes are available, in contrast to many other confinement hosts, where the pore dimensions are distributed. The available pore sizes include the region of 1–2 nm, which is not well covered by other materials.[30] In Ref. 25 we used three MFU-type MOFs (MFU stands for "Metal-Organic Framework Ulm-University") as host materials with pore sizes between 1.19 and 1.86 nm to study the dynamics of the glass former glycerol in confined geometry.[25] In small pores (MFU-4), no bulk dynamics could be observed, indicating that the correlation length in glycerol is larger than the pore diameter of 1.19 nm, even at the highest investigated temperature of 380 K. While for the MOFs with bigger pores the expected confinement effects were partly observed, the interpretation of these results was complicated by the presence of two types of pores in the structure with 1.2 and 1.8 nm (MFU-4l) and by a partial interdigitation of the pore framework (MFU-1). Nevertheless, these earlier investigations have demonstrated that MOFs indeed seem to be ideal host systems for confinement measurements, with weak molecule pore-wall interactions and essentially three-dimensional pores with sizes that are comparable to the cooperativity length scales in typical glass-forming liquids.

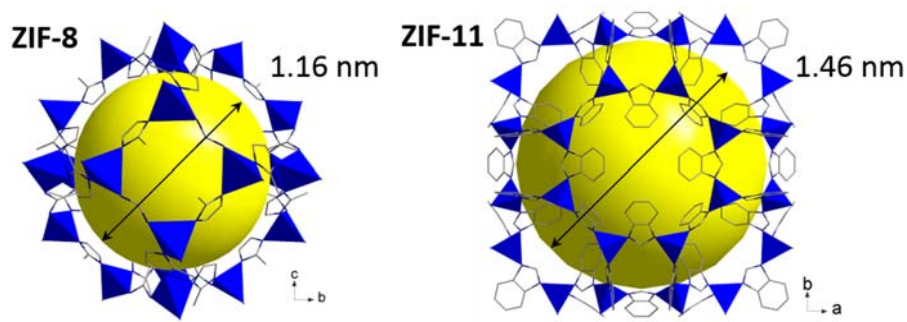

FIG. 1. Crystal structures of the ZIFs used in this study; both are drawn to a common scale. Zn ions are represented by polyhedra and the pores by spheres. H atoms are omitted for clarity.

For the present work, we have selected two special MOFs, namely the microporous zeolitic imidazolate frameworks (ZIFs) ZIF-8 and ZIF-11 with two different, well-defined pore sizes (Fig. 1). ZIFs have zeolite-like topologies.[31,32,33] They are composed of tetrahedrally-coordinated transition metal ions [typically Zn(II) or Co(II)] connected by imidazolate linkers. The metal-imidazole-metal link has a large angle of ca. 145° which is analogues to the 145° Si-O-Si angle found in zeolites, and thus ZIFs adopt the same framework topologies as zeolites. Additionally, they are known as porous materials with exceptionally high chemical and thermal stability and with strongly hydrophobic internal surfaces, making the formation of surface layers of the guest material very unlikely. ZIF-8 and ZIF-11 crystallize in a cubic crystal system, in the space group I$\bar{4}$3m and Pm$\bar{3}$m, respectively. An important structural feature of these ZIFs is, that they possess rather large pores (ZIF-8: 1.16 nm in diameter, ZIF-11: 1.46 nm), connected by significantly smaller pore apertures (ZIF-8: 0.34 nm, ZIF-11: 0.30 nm). Therefore, these materials exhibit well-defined, independent cavities and we do not assume significant interactions between molecules in different pores. These structural properties (summarized in Table S1 in the supplementary material[34]) make them attractive host materials for studying dynamics of confined liquids.

In the present work, we investigate the molecular dynamics of supercooled glycerol confined in these hosts via broadband dielectric spectroscopy. We compare our results to previous studies of the cooperativity length scale in this system. Combining the results from confinement measurements of glycerol with the relative temperature dependence of its amorphous-order length scale as recently deduced from nonlinear dielectric spectroscopy[11,12] provides interesting hints at the temperature evolution of cooperativity in this prototypical glass former.

## II. EXPERIMENTAL DETAILS

### A. Preparation and characterization of glycerol loaded ZIFs

All reagents were used as received from commercial suppliers. ZIF-8 and ZIF-11 were synthesized according to the previously described procedures.[35,36] The two ZIFs used in this work consist of the same metal ion, Zn(II), but differ in the organic linkers. ZIF-8 was synthesized using 2-methylimidazole (H-MeIM),[35] whereas benzimidazole (H-PhIm) was used in the preparation of ZIF-11.[36] The phase purity was confirmed by X-ray powder diffraction (XRPD) measurements (Figs. S1 and S2 in the supplementary material[34]). The XRPD data were collected in the 2θ range of 4–70° with 0.02° steps using a Seifert XRD 3003 TT diffractometer equipped with a Meteor 1D detector. To load glycerol into the pores, ZIF samples (30 mg) were degassed for 20 h at 120 °C in vacuum and then placed in an open vial into a Schlenk tube containing glycerol (2 mL). The Schlenk tube was heated for 24 h at 85 °C in vacuum (approx. 1 mbar). The amount of glycerol loaded into the pores was determined by thermogravimetric (TG) analysis (Table I and Fig. S3 in the supplementary material[34]). The TG analysis was performed with a TA Instruments Q500 analyzer in the temperature range of 25–700 °C under nitrogen atmosphere at a heating rate of 5 K min$^{-1}$. Models of unit cells of the ZIFs loaded with glycerol (shown in Fig. S4 and S5 in the supplementary material[34]) were created by the "Sorption Tools" module of Accelrys Materials Studio 2017, employing a Metropolis sampling scheme to find appropriate positions of the glycerol molecules (loading at 298 K to a fixed target pressure of 100 kPa). During the sampling, all framework lattice atoms were fixed at their crystallographic positions. The number of glycerol molecules per unit cell obtained from sorption simulations is given in Table I.

### B. Broadband dielectric spectroscopy

Broadband dielectric spectra of the dielectric loss of the samples were collected in the frequency range of 1 mHz - 3 GHz combining two experimental techniques.[37] For the low-frequency range (1 mHz – 1 MHz), a frequency response analyzer (Novocontrol $\alpha$-Analyzer) was used. For the radio frequency and microwave region (1 MHz – 3 GHz), two impedance analyzers using a coaxial reflectometric setup were employed. These measurements were performed by an Agilent 4294A impedance analyzer for frequencies below 100 MHz and by an Agilent 4991B for higher ones. In this technique, the sample capacitor is mounted at the end of a specially designed coaxial line.[38] Cooling and heating of the samples was achieved by a closed-cycle refrigerator and a nitrogen-gas cryostat.

All measurements were carried out on powder samples filled in parallel-plate capacitors to avoid pressure-induced deterioration of the glycerol-loaded pores that could occur when pressing pellets.

For the capacitors, a plate distance of approximately 100 µm was realized. To minimize voids in the samples, slight pressure was applied to the filled capacitors. After mounting the loaded capacitors into the cooling devices, they were kept under vacuum at room temperature for at least 24 h before starting the dielectric measurements. This procedure ensured that all adsorbed contaminations (mainly water) were removed from the sample surface.

## III. RESULTS AND DISCUSSION

### A. Sample characterization

Glycerol was introduced to the pores via vapor diffusion from the gas phase. To confirm that the glycerol loading did not influence the MOF crystal structure, XRPD measurement before and after the glycerol loading were carried out (Figs. S1 and S2 in the supplementary material[34]). The glycerol content in the loaded samples was determined by TG analysis. The TG curves (Fig. S3 in the supplementary material[34]) reveal that the glycerol release occurred between 90-210 °C with the mass loss -27.4 % for ZIF-8 and -22.3 % for ZIF-11. As shown in Table I, the experimental loading of glycerol per unit cell of ZIF-8 and 11 closely matches the theoretical values obtained from sorption simulations.

TABLE I. Overview of simulated and experimentally found amount of glycerol molecules per unit cell of ZIF-8 and ZIF-11.

| ZIF-n (unit cell composition) | ZIF-8 ($C_{96}H_{120}N_{48}Zn_{12}$) | ZIF-11 ($C_{672}H_{480}N_{192}Zn_{48}$) |
|---|---|---|
| Experimentally found number of glycerol molecules per unit cell/pore | 11.4/5.7 | 44.2/22.1 |
| Simulated number of glycerol molecules per unit cell (average) | 11.2 | 41.8 |
| Volume ratio of pores to host material ($\Phi_f$)[a] | 44% | 38% |
| Pore diameter (nm) | 1.16 | 1.46 |

[a] From Ref. 39.

### B. Broadband dielectric spectroscopy

As an example for our broadband dielectric-spectroscopy results, Fig. 2 shows the dielectric loss, the imaginary part of the permittivity $\varepsilon''$, of glycerol in ZIF-11 as a function of frequency for various temperatures. At first glance, two relaxation processes can be immediately identified, leading to the typical peaks in the dielectric-loss spectra, which shift to lower frequencies upon cooling. (In the following, the faster one will be denoted as process 1 and the slower one as process 2.) A comparison of process 1 with the $\alpha$ relaxation in bulk glycerol,[40] i.e., the main reorientational motion of the molecules, which is closely coupled to the glass transition,[41] reveals similar loss-peak frequencies. Therefore, we ascribe the fastest process detected in the spectra to the $\alpha$ relaxation of glass-forming glycerol confined in ZIF-11. At the lowest frequencies and high temperatures, an additional increase in $\varepsilon''$ for decreasing frequency is observed (e.g., below about 1 Hz for the 298 K spectrum). This indicates another contribution (termed process 3), which is only partly resolved due to the low-frequency limit of the measured spectra. Consequently, the spectra of Fig. 2 are fitted by the sum of three relaxation functions. Relaxations 1 and 2 are described by a Havriliak-Negami (HN) function, whereas for process 3 the Cole-Cole (CC) equation was used, the latter being a special case of the HN function with symmetrical broadening.[42] Both are empirical functions and often employed to parameterize dielectric relaxation processes. Separate measurements of the empty ZIF-11 at room temperature (not shown) did not exhibit any significant indications of relaxational dynamics. Therefore, we can exclude that residual amounts of solvent or of water within the sample[43] contribute to the detected relaxational response of glycerol enclosed in ZIF-11.

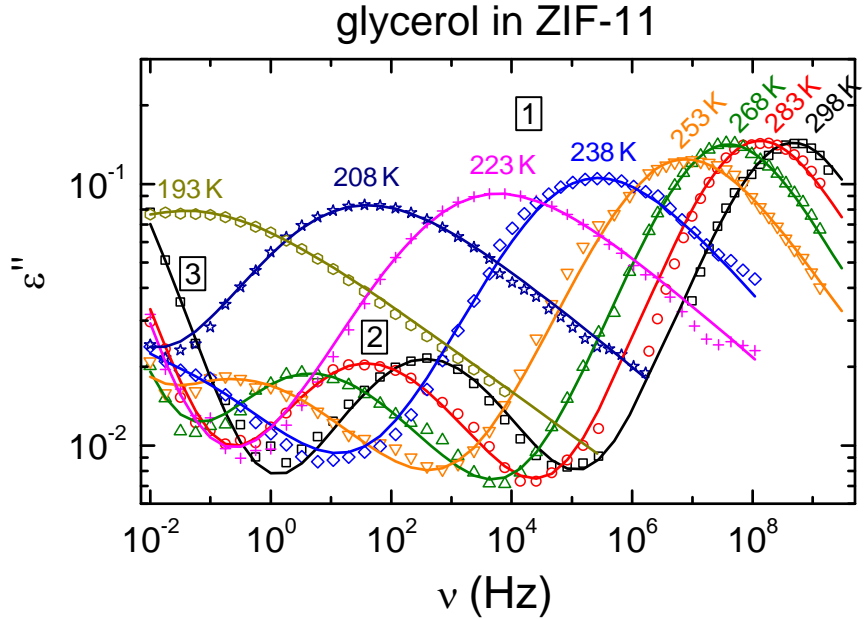

FIG. 2. Frequency dependence of the dielectric loss of glycerol confined in ZIF-11 for different temperatures. The lines through the points are fits by the sum of two HN functions for the two well-pronounced relaxations and a CC function to formally account for the increase of the loss found at low frequencies and high temperatures. The numbers denote the three contributions to the spectra.

What is the origin of the two low-frequency processes observed in the spectra of Fig. 2? Ionic impurities in the sample that are sufficiently small to pass through the pore apertures, can lead to ionic charge transport, giving rise to dc and ac contributions to the complex conductivity $\sigma^* = \sigma' + i\sigma''$.[44,45] As the dielectric loss and the real part of the conductivity are related via $\varepsilon'' \propto \sigma'/\nu$, an increase of $\varepsilon''$ towards low frequencies will result as is also observed in bulk glycerol.[46] The slowest process (3), which in the spectra of Fig. 2 only appears as an additional low-frequency increase for the highest temperatures ($T \geq 253$ K), can probably be ascribed to such conductivity contributions. Therefore, it will not be further discussed in the following and the CC function used to fit this spectral feature should be regarded as a purely phenomenological description.

Concerning process 2, two different explanations can be considered:[14,15,16,47,48,49] In confined dipolar systems, in addition to the $\alpha$ relaxation arising from the relatively freely reorienting guest molecules, a layer of molecules with slowed down dynamics due to interactions with the pore walls can exist, leading to a separate loss peak at much smaller frequency.[14,48,49] However, one should be aware that the inner pore walls of the ZIFs are strongly hydrophobic making strong interactions with the dipolar glycerol molecules unlikely. Alternatively, another process can arise from the heterogeneous nature of the guest-host system: It was shown long ago by Maxwell and Wagner that samples composed of two different dielectric materials can exhibit a nonintrinsic relaxation-like process in the dielectric spectra, nowadays termed Maxwell-Wagner (MW) relaxation.[50,51,52] In principle, based on the known dielectric properties of the bulk and host material, theoretical spectra that are modified by the MW effect may be calculated using the following relation:[47,52]

$$\varepsilon_c(\nu) = \varepsilon_m(\nu) \frac{n\varepsilon_f(\nu)+(1-n)\varepsilon_m(\nu)+\Phi_f(1-n)(\varepsilon_f(\nu)-\varepsilon_m(\nu))}{n\varepsilon_f(\nu)+(1-n)\varepsilon_m(\nu)-\Phi_f n(\varepsilon_f(\nu)-\varepsilon_m(\nu))} \qquad (1)$$

Here $\varepsilon_c(\nu)$ is the permittivity of the glycerol-loaded ZIFs, $\varepsilon_f(\nu)$ the permittivity of bulk glycerol, $\varepsilon_m(\nu)$ the permittivity of the empty ZIFs, $n$ is a factor referring to the geometry of the pores (for spherical cages it is 1/3), and $\Phi_f$ is the volume ratio of the confined material and the host system (Table I). Calculating $\varepsilon_c(\nu)$ using Eq. (1), leads to theoretical loss spectra that exhibit two relaxation peaks. (Variations of $\Phi_f$ between 38 and 44% only cause a small change of $\tau$; therefore this calculation can be taken representatively for both ZIF systems.) One of them arises from the $\alpha$ relaxation, which, due to the MW effects, can be shifted towards somewhat higher frequencies compared to the bulk.[47] The second one reflects the non-intrinsic, heterogeneity-related MW relaxation mentioned above. The resulting peak positions of the latter do not agree with those of process 2. However, one cannot exclude that the assumptions of the used MW model are oversimplified. Overall, one has to state that the origin of the detected process 2 is not finally clarified yet, but the main interest of the present work lies in the investigation of the $\alpha$-relaxation dynamics of glycerol. Thus, in the following we concentrate on the confinement-induced behavior of process 1.

The dielectric spectra of glycerol confined in ZIF-8 also exhibit two relaxational processes, in addition to the conductivity contribution showing up at low frequencies and high temperatures (see Fig. S6 in the supplementary material[34]). Processes 1 and 2 both are present, occur in a similar frequency region as for glycerol in ZIF-11, and can be rationalized in a similar way. Especially, relaxation 1 in ZIF-8 again corresponds to the $\alpha$ relaxation of glycerol. In contrast to ZIF-11, the symmetrical CC function was sufficient to fit process 2 of glycerol in ZIF-8.

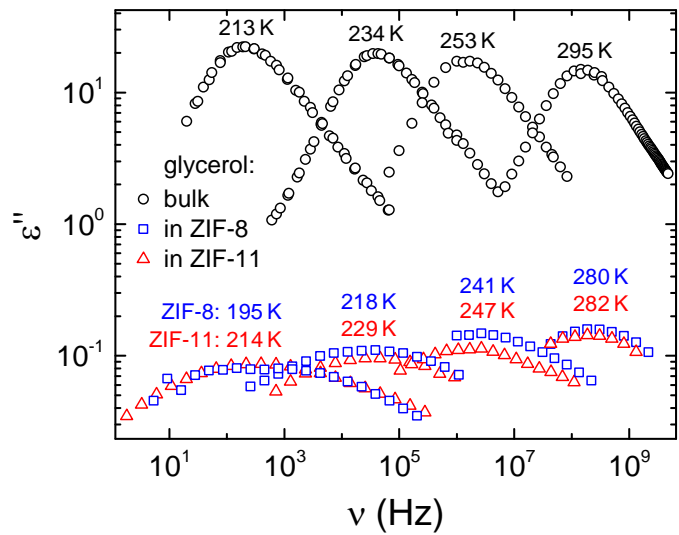

FIG. 3. Frequency dependence of the dielectric loss of glycerol in the $\alpha$-peak region (process 1) confined in ZIF-11 (triangles) and ZIF-8 (squares), and of bulk glycerol (circles).[40] The temperatures of the confined systems were selected to achieve an approximate match of the peak frequencies of the bulk results.

Figure 3 shows typical loss spectra in the $\alpha$-relaxation region (process 1) of glycerol confined in ZIF-11 and ZIF-8, compared to those of the bulk, measured at four different temperatures.[40] For the confined samples, the temperatures of the shown spectra were selected to achieve an approximate match of the peak frequencies with those of the bulk spectra. The first insight gained from this plot is the enormous reduction of the relaxation strength $\Delta\varepsilon$ induced by the confinement. The performed fits suggest a reduction of $\Delta\varepsilon$ of about two orders of magnitude. This phenomenon is a common finding for confined systems.[15,16,19,47] Obvious reasons for this behavior are the decreased amount of supercooled glycerol per volume in the confined samples (because part of the volume is occupied by the host material) and the incomplete space filling of the sample powder. However, when considering the volume ratio of the guest-host system of around 38 to 44%, it becomes clear that there has to be another contribution to the found strong reduction of $\Delta\varepsilon$. In fact, as pointed out, e.g., by Richert,[47] the relaxation strengths of heterogeneous mixtures of different dielectrics do not add up in a simple way. The difference in $\Delta\varepsilon$ between the two confined systems is rather marginal and probably due to the different combinations of the aforementioned effects in these systems.

Figure 3 also reveals that the $\alpha$ relaxations of the confined glycerol are considerably broadened compared to the bulk. At room temperature, the bulk loss-peak features a half width of 1.6 decades, already somewhat smaller than the values of 2.2 in ZIF-11 and 2.1 in ZIF-8. When approaching 200 K, the half width for the bulk only increases to 2.1 decades, whereas in ZIF-11 it even becomes 4.9 and in ZIF-8 3.7 decades. Such a strong broadening is a common finding for confinement measurements and may be explained by a distribution of relaxation times due to interactions of the glycerol molecules with the pore walls and/or a variation in the number of glycerol molecules per pore.[14,15,19,22,25]

Comparing the temperatures in Fig. 3 already provides first hints at the significantly different slowing down of the molecular dynamics in the two confined systems and bulk glycerol: For both confined systems, more or less strong deviations from the bulk temperatures show up implying the modification of molecular dynamics by the confinement. The dynamics of the molecular motions is characterized by the average relaxation time $\langle\tau\rangle$, which is related to the peak frequency via $\langle\tau\rangle \approx 1/(2\pi\nu_p)$. Thus Fig. 3 indicates a confinement-induced shift of the relaxation times of the guest material. This phenomenon certainly is the most interesting outcome of confinement measurements on supercooled liquids. As discussed in the introduction, ideally it should provide insights into the cooperativity length scales in glass forming systems.[15,16,17,18,22,24,25]

## C. Relaxation times

For a better comparison of the relaxation-time dynamics, $\langle\tau\rangle(1/T)$ is plotted in Fig. 4 for the two confined systems. The shown relaxation times represent the inverse circular peak frequency, $\langle\tau\rangle \approx 1/(2\pi\nu_p)$, calculated[53] from the fit parameters used to describe the dielectric spectra with the HN function (e.g., lines in Figs. 2), which is a good approximation of $\langle\tau\rangle$. For the CC function, used for relaxation 2 in ZIF-11, due to its symmetric shape $\langle\tau\rangle$ is equal to $\tau_{CC}$ as directly obtained from the fits. In Fig. 4, the present $\langle\tau\rangle(T)$ data are shown together with those for bulk glycerol published earlier (dashed lines).[54]

The relaxation times of process 2 (squares in Fig. 4) exhibit a linear Arrhenius-like behavior, with small deviations for ZIF-8. They are of comparable order of magnitude for both confined systems. Moreover, for glycerol in MFUs a similar process was found just in this region.[25] Thus, it is likely that these relaxations are all of the same origin. As mentioned above, they may be of Maxwell-Wagner origin or arise from layers of glycerol molecules interacting with the pore walls, the latter being unlikely due to the hydrophobic internal surfaces of the ZIFs. In the context of the present work, aiming at the clarification of cooperativity effects of the $\alpha$ relaxation, these slow processes are of minor relevance.

The dashed lines in Fig. 4 indicate the temperature-dependence of the $\alpha$-relaxation time of bulk glycerol.[54] It deviates from thermally activated Arrhenius behavior, which is a typical feature of glass-forming liquids and can be described by the empirical Vogel-Fulcher-Tammann (VFT) law:[55,56,57,58]

$$\tau = \tau_0 \exp\left(\frac{DT_{\text{VF}}}{T-T_{\text{VF}}}\right) \tag{2}$$

(In fact, for better readability of the figure the dashed lines show the VFT fit curve instead of the actual experimental $\tau$ values.[54]) In this equation, $\tau_0$ corresponds to an inverse attempt frequency, $T_{\text{VF}}$ is the Vogel-Fulcher temperature marking a divergence of $\tau$ below the glass transition, and $D$ represents the so-called strength parameter, which can be used to quantify the degree of deviation from Arrhenius behavior.[58] For $T_{\text{VF}} = 0$, Eq. (2) corresponds to Arrhenius behavior.

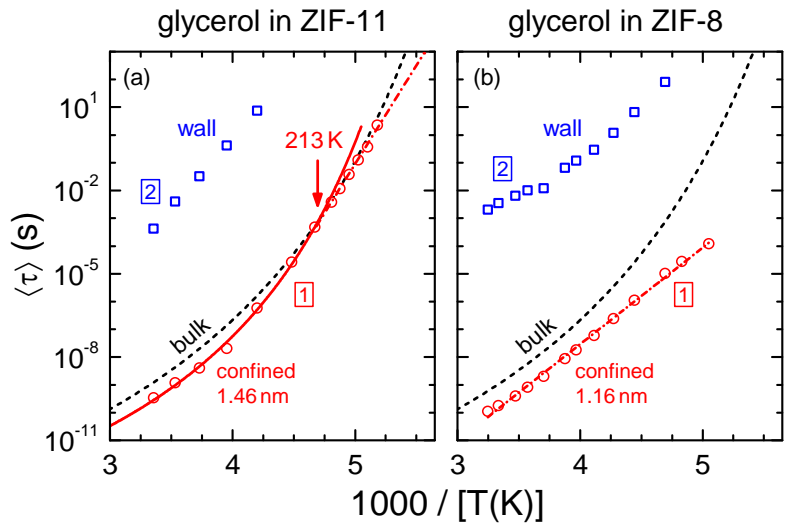

FIG. 4. Arrhenius plot of the average relaxation times of glycerol confined in ZIF-11 (a) and ZIF-8 (b). The framed numbers denote the two detected relaxation processes. For comparison, the dashed line in each frame represents the VFT fit of $\langle\tau\rangle$ of bulk glycerol as reported in Ref. 54. The circles indicate the $\alpha$-relaxation times (process 1), the squares those of the slow process 2. The solid line in (a) is a fit of the high-temperature part of the $\alpha$-relaxation-time curve with the VFT function, Eq. (2). The dash-dotted lines are linear fits, indicating Arrhenius behavior. The arrow in (a) indicates the transitions from VFT to Arrhenius behavior. The parameters of the VFT fits are: $\tau_0 = 3.9 \times 10^{-15}$ s, $D = 16$, $T_{VF} = 132$ K (bulk[54]); $\tau_0 = 4.0 \times 10^{-15}$ s, $D = 11$, $T_{VF} = 149$ K (ZIF-11). The energy barriers $E_a$ for the Arrhenius part of the $\alpha$ relaxation are: $E_a = 1.4$ eV (ZIF-11) and $E_a = 0.69$ eV (ZIF-8).

The circles in Fig. 4(a) show the relaxation times for glycerol in ZIF-11. At temperatures above about 213 K, the experimental data can be well fitted by a VFT-law (solid line). However, below this temperature $\tau(T)$ is better described by the Arrhenius function (dash-dotted line). As mentioned in the introduction, such a transition from VFT to Arrhenius behavior at low temperatures in confinement points to growing cooperativity of molecular dynamics under cooling beyond the pore diameter.[15] Within this framework, the non-canonical slowing down of molecular motion when approaching the glass temperature is explained by an increasing number of cooperatively moving molecules. This implies a growth of the correlation length scale $L_{\text{corr}}$, characterizing the size of the cooperatively rearranging regions, consistent with a phase-transition-related origin of the glass transition.[5,6,12,9] The increasing cooperativity leads to a growing effective energy barrier with decreasing temperature and explains the commonly observed super-Arrhenius behavior of $\tau(T)$. In bulk materials, $L_{corr}$ can get larger without restrictions, but in confinement the pore diameter of the host system limits the maximum correlation length.[15] Within this scenario, the observed crossover in $\tau(T)$ of ZIF-11 can be explained by $L_{corr}(T)$ of bulk glycerol exceeding the pore diameter of 1.46 nm below about 213 K. In the confined system, the cooperativity length cannot grow beyond 1.46 nm and remains temperature independent and of the order of the pore size under further cooling. Therefore the energy barrier does not increase anymore and Arrhenius temperature dependence of $\tau(T)$ is observed. It should be noted that an alternative explanation of such a transition from VFT to Arrhenius in confinement was recently suggested, based on the assumption of a separate glass transition of molecules strongly interacting with the pore walls.[59] However, such an explanation seems unlikely in the present case, because in ZIFs, in contrast to most other host materials, only weak molecule pore-wall interactions are expected.

At high temperatures, the deduced $\tau(T)$ of glycerol confined in ZIF-11 is by about a factor of 3-4 smaller than for the bulk material [Fig. 4(a)]. As mentioned above and thoroughly discussed in a review article by Richert[47], such an acceleration of the $\alpha$ relaxation in a confined system may rather trivially arise from the heterogeneity of the guest-host system, not being related to the "real" confinement effects arising from the length scales involved in the glass transition. When using Eq. (1) to estimate the shift of the relaxation times expected within this scenario, we find that it indeed is consistent with the experimental observations at high temperatures ($1000/T < 4$ K$^{-1}$). However, it should be noted that $\tau(T)$ of the confined sample exhibits a somewhat stronger overall temperature dependence than the bulk (at least at $T > 213$ K), which leads to an approach of the bulk $\tau(1/T)$-curve in Fig. 4(a) at low temperatures. This cannot be explained by the mentioned heterogeneity effects and demonstrates some confinement-induced variation of the temperature-dependent cooperativity of the bulk-like $\alpha$-relaxation dynamics within the pores. It should be noted that qualitatively similar behavior was previously also reported by Fischer et al. for glycerol in a different MOF (MFU-4l).[25] Anyway, for glycerol confined in ZIF-11 there is a clear transition of $\tau(T)$ from VFT to Arrhenius behavior at low temperatures, just as expected within the framework of a cooperativity-driven glass transition. Notably, until now such a temperature-dependent crossover, triggered by a confinement-induced suppression of a further growth of cooperativity length, was only rarely observed.[15,19,25] Instead, very often the whole $\tau(T)$ curves in confined systems were found to be completely shifted compared to the bulk, allowing, at best, an estimate of a lower limit of $L_{corr}$ only.

The latter scenario is also found for ZIF-8 [Fig. 4(b)], where the temperature-dependent $\alpha$-relaxation time (circles) exhibits a clear shift to faster relaxations rates compared to the bulk and Arrhenius behavior in the entire investigated temperature range. At the lowest temperature of 198 K, the deviation from the bulk $\tau$ exceeds three decades. This finding can be well understood by the strong confinement in the small ZIF-8 pores with diameters of 1.16 nm. Obviously, even at the highest measured temperature of 308 K, the correlation length of bulk glycerol still is larger than the pore diameter of ZIF-8. This prevents a temperature-dependent variation of $L_{corr}$ within the small pores of ZIF-8. Consequently, for the confined system there is no indication of the typical super-Arrhenius temperature dependence of $\tau(T)$ of glass-forming systems in the whole temperature range. This finding is well consistent with those for glycerol in MFU-4 with pore sizes of 1.19 nm, whose $\tau(1/T)$ curves are much closer to Arrhenius behavior than bulk glycerol.[25] Overall, the present results on glycerol in ZIF-8 provide a lower limit of 1.16 nm for the cooperativity length scale in glycerol at temperatures up to 308 K.

### D. Cooperativity length

Figure 5 summarizes the findings concerning the temperature dependence of the cooperativity length scale in glycerol, based on the present work and literature data. To our knowledge, absolute values for $L_{corr}$, deduced from experimental data, until now were only provided by the dynamic calorimetry investigation of Hempel et al.[60] (stars in Fig. 5) and by the present measurements in ZIF-11 (1.46 nm at 213 K; triangle), the latter revealing a clear crossover from cooperativity-dominated

VFT to non-cooperative Arrhenius behavior. Based on multidimensional nuclear-magnetic-resonance experiments,[61] the length scale $\xi_{het}$ of dynamic heterogeneities in glycerol at temperatures around 200 K was found to be about 1 - 1.3 nm,[62] of similar order as $L_{corr}$ determined in the present work. However, as noted, e.g., in Ref. 61, $L_{corr}$ and $\xi_{het}$ do not necessarily have to agree.

In Ref. 11, in agreement with an earlier work using similar methods,[10] information on the temperature dependence of the number of correlated molecules, $N_{corr}$, was obtained, based on measurements of the third-order harmonic component of the dielectric susceptibility.[63] From the definition of $N_{corr}$, it immediately follows that $L_{corr} \propto (N_{corr})^{1/3}$. While no absolute values of $N_{corr}$ (and thus of $L_{corr}$) could be obtained by this approach, the value at 213 K from the present work (triangle in Fig. 5) can be used to scale these data as shown by the plusses in Fig. 5. For both ZIF-8 and MFU-4,[25] no super-Arrhenius behavior was found in the whole investigated temperature range. Thus, as discussed above, based on these measurements only lower limits of $L_{corr}$ can be provided. The temperatures of the corresponding data points shown in Fig. 5 represent the maximum T covered by these experiments (only above this temperature, $L_{corr}$ may decrease below this limit). Moreover, from the results on glycerol confined in MFU-4l, reported in Ref. 25, an upper limit of $L_{corr}$ can be derived. This is based on the fact that for this confined system super-Arrhenius behavior of $\tau(T)$ was found in the entire investigated temperature range, i.e., $L_{corr}$ of glycerol obviously did not exceed the pore size of 1.86 nm, even at the lowest covered temperature of 205 K. The corresponding data point (circle in Fig. 5) is shown at the lowest investigated temperature of this study. A similar situation was found for glycerol in nanoporous sol-gel glass with a pore size of 2.5 nm, reported by Arndt et al.,[49] which is shown by the "x" in Fig. 5. The shaded areas in Fig. 5 indicate the excluded regions of $L_{corr}$, as inferred from the shown lower and upper limiting values. Results on glycerol confined in MFU-1, reported in Ref. 25, are not presented in Fig. 5 as their interpretation is hampered by partial interdigitation of the host framework, leading to a distribution of pore sizes.[25]

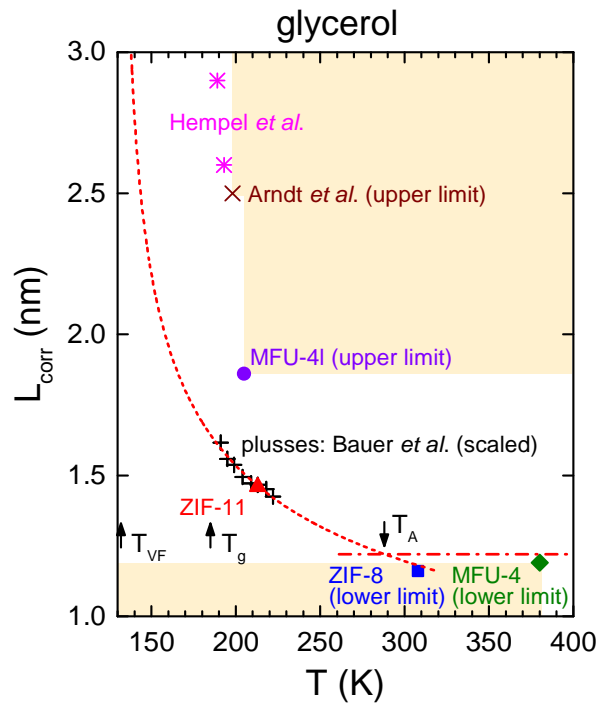

FIG. 5. Temperature dependence of the cooperativity length $L_{corr}$ of glycerol as deduced from the present work and literature data. The closed symbols show the results from confinement measurements performed in our group using ZIF- (present work) and MFU-type MOFs[25] as host materials. For ZIF-8 and MFU-4, only lower limits of $L_{corr}$ can be provided and for MFU-4l an upper limit was reported. The shaded areas indicate the excluded ranges for $L_{corr}$ inferred from these limiting values. Literature data from Arndt et al.[49] (confinement in nanoporous glass; upper limit only) and from Hempel et al.[60] (dynamic calorimetry; absolute values) are also included. The plusses show the temperature-dependent correlation length from measurements of the 3rd harmonic susceptiblity[11], scaled to the value of 1.46 nm at 213 K that was obtained from the present results on glycerol in ZIF-11 [cf. Fig. 4(a)]. The dashed line indicates a fit of $L_{corr}(T)$ from Ref. 11 (plusses) with a critical law, $L_{corr} \propto (T-T_c)^{-\gamma}$, with $\gamma = 0.28$ and $T_c$ fixed to the Vogel-Fulcher temperature, i.e., $T_c = T_{VF} = 132$ K.[54] The dash-dotted line indicates a possible temperature-independent $L_{corr}$ above $T_A \approx 288$ K, where $\tau(T)$ was reported to cross over from VFT to Arrhenius behavior.[67,71] The arrows indicate the Vogel-Fulcher temperature, glass temperature, and VFT-Arrhenius transition temperature (from left to right).

The dashed line in Fig. 5 represents a fit of the temperature-dependent correlation length from Ref. 11 (plusses) with a critical law, $L_{corr} \propto (T-T_c)^{-\gamma}$. Of course, the temperature range covered by this data set is rather restricted and, in principle, should not allow drawing meaningful conclusions on the involved parameters or on the development of $\tau(T)$ at lower and higher temperatures. However, within scenarios assuming a "hidden" phase transition as the true reason of the glass transition, this transition is expected to occur close to the Kauzmann[64] or the Vogel-Fulcher temperature (both are usually of similar magnitude[65]), where $\tau(T)$ should diverge [cf. Eq. (2)]. Therefore, for the fit of $L_{corr}(T)$ we could fix $T_c$ at $T_{VF} = 132$ K[54] which strongly enhances the reliability of the obtained fit curve and fit parameters. With a critical exponent $\gamma = 0.28$, resulting from the fit, the experimental data can be reasonably described in this way (dashed line in Fig. 5).[66] However, at temperatures above about 300 K, the extrapolated fit curve falls below the lower limit of $L_{corr}$ of about 1.2 nm, derived from our confinement investigations of glycerol in ZIF-8 and MFU-4 (lower shaded region in Fig. 5). To explain this apparent discrepancy, one should be aware that for various glass-forming materials, a transition of $\tau(T)$ from VFT to Arrhenius behavior was proposed to occur above a temperature $T_A$.[67,68,69,70] For glycerol, such a crossover was reported to arise at a temperature $T_A$ of about 290 K.[67,71] When assuming that the VFT temperature dependence is caused by an increase of the effective energy barrier at low temperatures due to increasing cooperativity,[2,5,6,7] the proposed high-temperature Arrhenius behavior above $T_A$, corresponding to a constant energy barrier, would imply constant (or absent) cooperativity. Indeed, direct interactions with the nearest-neighbor shell should be possible at all temperatures and, in first respect, constitute Arrhenius-type relaxation dynamics. Such a scenario is well consistent with the results of Fig. 5 on $L_{corr}(T)$ when assuming a crossover from the critical law (dashed line) to constant behavior (dash-dotted line) close to $T_A$. When adjusting the level of the latter to obtain a crossover between critical and temperature-independent $L_{corr}(T)$ just at $T_A = 288$ K,[67] we arrive at $L_{corr} \approx 1.22$ nm for $T > T_A$. This value is well consistent with the lower limits for this length scale provided by our confinement measurements.

At low temperatures, while being consistent with the upper-limit values provided by the MFU-4l[25] and sol-gel-glass confinement investigations,[49] the extrapolated critical law for $L_{corr}(T)$ with $T_c = T_{VF}$ does not match the two data points from Ref. 60 (stars in Fig. 5), based on dynamic calorimetry. It is clear that the temperature range of the fitted third-order dielectric data (plusses) is too restricted to allow for meaningful fits without the critical temperature being fixed at $T_{VF}$. In any case, to account for the data by Hempel et al.,[60] $T_c$ would have to be much larger than $T_{VF}$ and a divergence of $L_{corr}$ close to $T_g$ of about 185 - 189 K,[40,60] would have to be assumed. This seems quite unrealistic because $T_g$ only marks a purely dynamical effect and the possible phase transition into a state with "amorphous order" that may underlie the glass transition is expected at much lower temperature.[5,6,9]

## IV. SUMMARY AND CONCLUSIONS

In the present work, we have reported dielectric confinement measurements on glycerol in two different zeolitic imidazolate frameworks with well-defined pore sizes of 1.16 and 1.46 nm. Our findings demonstrate that this class of MOFs can provide well-suited host materials for confinement investigations of glass-forming liquids. In both investigated systems, the $\alpha$ relaxation of confined glycerol could be clearly identified. It exhibits the broadening and amplitude reduction as known from other confined systems. Most importantly, depending on temperature and pore size its relaxation time reveals different degrees of deviation from that of the bulk. In contrast to the previous investigation of glycerol confined in MOFs,[25] the ZIFs used here do not suffer from the presence of two types of pores, interdigitation, or partial pore filling. Thus, our present findings allow for significant conclusions concerning the cooperativity length scale $L_{corr}$ of glassy dynamics in glycerol: The measurements of glycerol in ZIF-8 reveal a lower limit $L_{corr}(T) > 1.16$ nm for $T < 308$ K while those for the ZIF-11 host material provide an absolute value of 1.46 nm at 213 K. These values are well consistent with earlier estimates of upper or lower limits $L_{corr}$ at different temperatures, based on confinement investigations of glycerol.[25,49] All these results can be combined with the information on the temperature dependence of $L_{corr}$, recently derived from nonlinear dielectric measurements.[11] We find that the available data are in good accord with a critical increase of the cooperativity length scale in glassforming glycerol and a divergence close to the Vogel-Fulcher temperature. This supports the notion that there is a growth of amorphous order when approaching the glass transition, which is due to an underlying phase transition arising significantly below $T_g$. However, one should be aware that the overall variation of the deduced length scale between the low-viscosity liquid at high temperatures and the solid glass somewhat below $T_g$ is rather limited. Only at temperatures significantly below the glass temperature, $L_{corr}$ varies considerably and finally diverges. However, in measurements with realistic cooling rates, of course this divergence does not come into effect because the system falls out of equilibrium at the glass transition.

Notably, the confinement data seem to indicate that at high temperatures, above about 290 K, $L_{corr}$ becomes temperature-independent and levels off at about 1.22 nm. This would be consistent with the often assumed crossover of the temperature-dependent relaxation time from VFT to Arrhenius behavior and certainly deserves further exploration. When considering the glycerol-molecule size of roughly 0.5 - 0.6 nm,[72,73] the deduced limiting high-temperature value of the correlation length of about 1.22 nm seems reasonable, essentially reflecting next-neighbor interactions only.

**SUPPLEMENTARY MATERIAL**
The supplementary material shows structural properties of the used ZIFs, models of unit cells of the ZIFs loaded with glycerol, TG-analysis results, and dielectric spectra of glycerol in ZIF-8.

**ACKNOWLEDGMENTS**
HB is grateful to the program "Chancengleichheit für Frauen in Forschung und Lehre" from the University of Augsburg for financial support via a fellowship.

# Glycerol confined in zeolitic imidazolate frameworks: The temperature-dependent cooperativity length scale of glassy freezing

## Supplementary material

M. Uhl,[1] J. K. H. Fischer,[1] P. Sippel,[1] H. Bunzen,[2] P. Lunkenheimer,[1] D. Volkmer,[2] and A. Loidl[1]

[1]Experimental Physics V, Center for Electronic Correlations and Magnetism, University of Augsburg, 86135 Augsburg, Germany
[2]Chair of Solid State and Materials Chemistry, Institute of Physics, University of Augsburg, 86135 Augsburg, Germany

TABLE S1. Structural characteristics of ZIF-8 and ZIF-11 determined by single crystal x-ray analysis.[1]

| ZIF-n | Pore aperture diameter (Å) | | | Pore diameter (Å) | Surface area (m$^2$/g) | Crystal system and space group |
|---|---|---|---|---|---|---|
| | 8-ring | 6-ring | 4-ring | | | |
| ZIF-8 | - | 3.4 | * | 11.6 | 1.947 | cubic (I$\bar{4}$3m) |
| ZIF-11 | 3.0 | 3.0 | * | 14.6 | 1.676 | cubic (Pm$\bar{3}$m) |

*The aperture sizes of the 4-rings in both ZIF-8 and -11 are negligible.

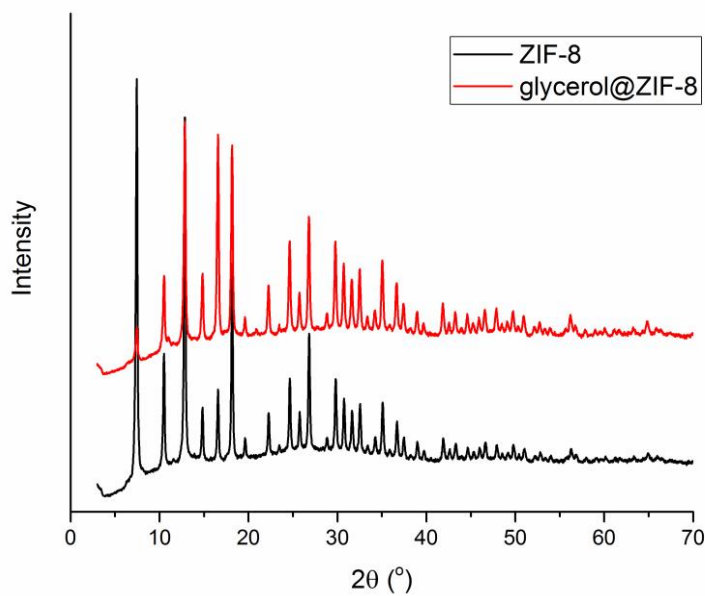

FIG. S1. XRPD patterns of an activated ZIF-8 sample and a glycerol-loaded ZIF-8 sample.

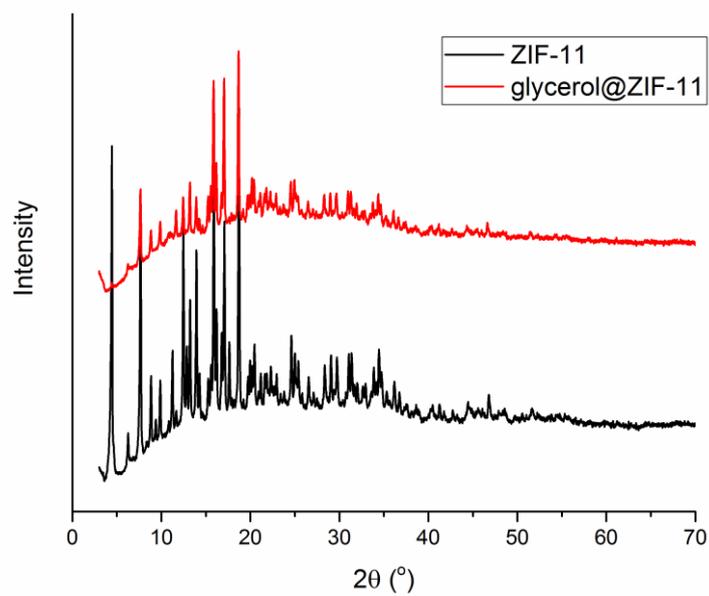

FIG. S2. XRPD patterns of an activated ZIF-11 sample and a glycerol-loaded ZIF-11 sample.

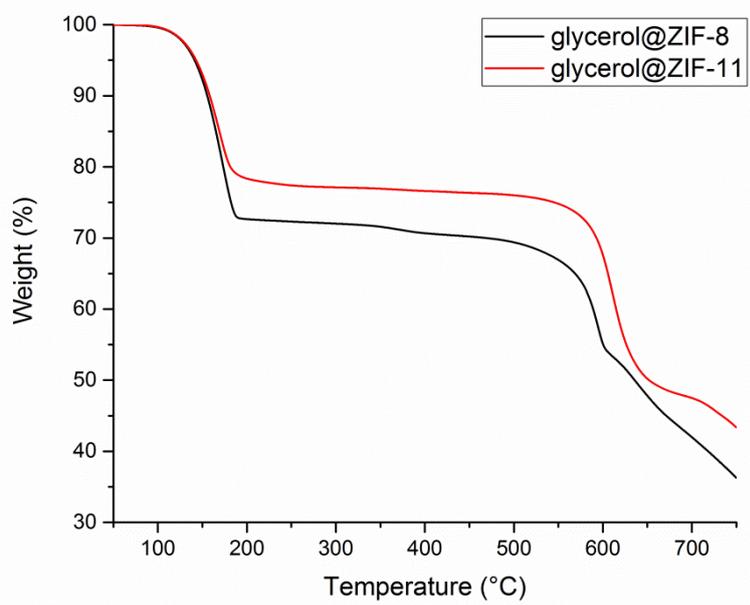

FIG. S3. Overlay of TG curves of ZIFs loaded with glycerol.

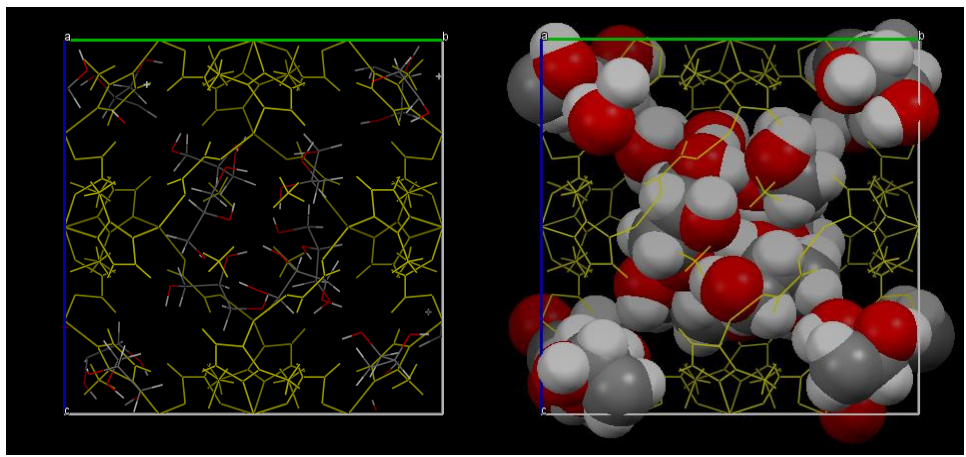

FIG. S4. A representative packing plot of wire model of ZIF-8 (yellow) viewed along the *a* axis, in which internal voids are filled with glycerol (12 molecules per unit cell, i.e. 6 molecules per pore).

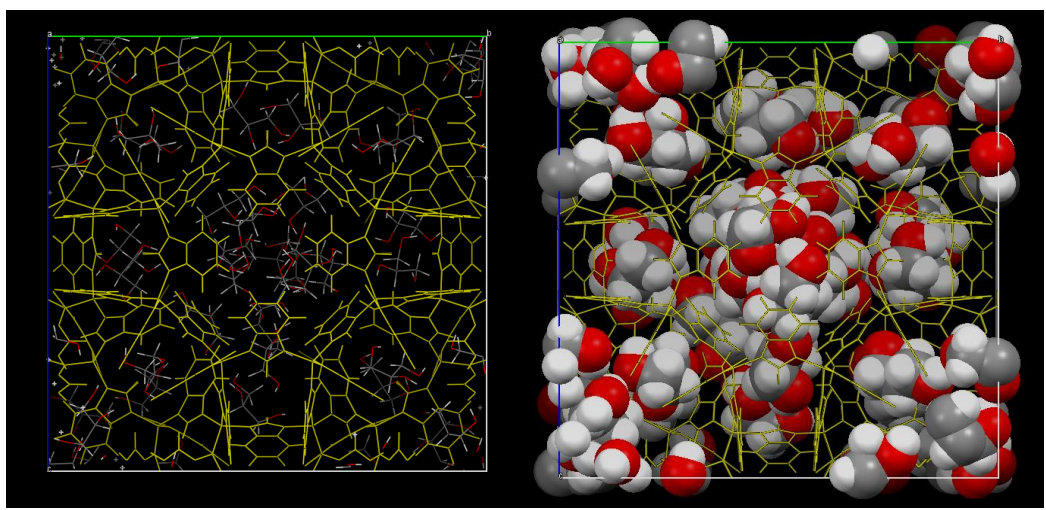

FIG. S5. A representative packing plot of wire model of ZIF-11 (yellow) viewed along the *a* axis, in which internal voids are filled with glycerol (42 molecules per unit cell, i.e. 21 molecules per pore).

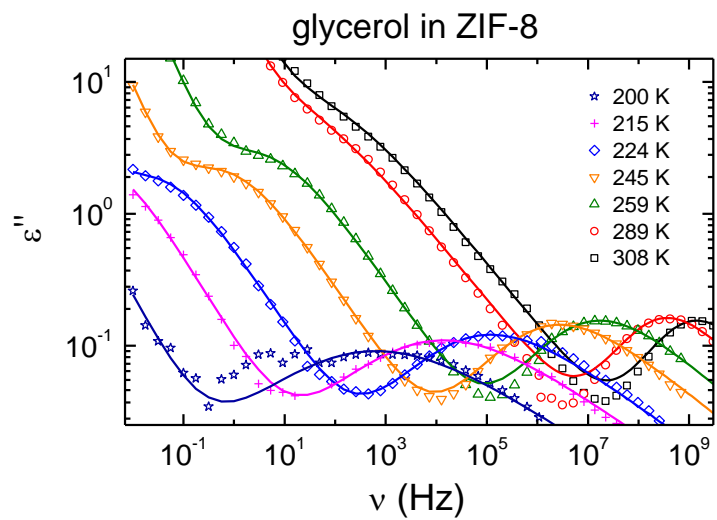

FIG. S6. Frequency dependence of the dielectric loss of glycerol confined in ZIF-8 for different temperatures. The lines through the points are fitted by the sum of two CC functions for the low-frequency effects and a HN function for the main relaxation.